\documentclass[aps,pra,12pt]{revtex4}

\usepackage{graphicx}  
\usepackage{amssymb}
\usepackage{hyperref}

\newcommand{\be}{\begin{equation}} 
\newcommand{\ee}{\end{equation}} 

\begin{document}

\title[Parametric Resonance in Bose-Einstein Condensates]{Parametric Resonance in Bose-Einstein Condensates with Periodic Modulation of Attractive Interaction}
\author{William Cairncross}
\affiliation{Institut f\"ur Theoretische Physik, Freie Universit\"at Berlin, Arnimallee 14, D-14195 Berlin, Germany}
\affiliation{Department of Physics, Queen's University at Kingston, Canada}
\affiliation{JILA, NIST and Department of Physics, University of Colorado, Boulder, CO 80309, USA}
\author{Axel Pelster}
\affiliation{Fachbereich Physik und Forschungszentrum OPTIMAS, 
	Technische Universit\"at Kaiserslautern, Germany}
\affiliation{Hanse-Wissenschaftskolleg, Lehmkuhlenbusch 4, 
	D-27733 Delmenhorst, Germany}


\begin{abstract}
We demonstrate parametric resonance in Bose-Einstein condensates (BECs) with attractive 
two-body interaction in a harmonic trap under parametric excitation by periodic 
modulation of the s-wave scattering length. We obtain nonlinear equations of motion for 
the widths of the condensate using a Gaussian variational ansatz for the Gross-Pitaevskii 
condensate wave function. We conduct both linear and nonlinear stability analyses for the equations of motion and find qualitative agreement, 
thus concluding that the stability of two equilibrium widths of a BEC might be inverted by parametric excitation.
\end{abstract}
\maketitle	
\section{Introduction}
The phenomenon of parametric resonance, when a system is parametrically excited and oscillates at one of its resonant frequencies,
is ubiquitous in physics: the phenomenon is found from simple classical 
systems such as the swing set and the vertically driven pendulum \cite{feynman}, to the Paul ion trap \cite{paul} and aspects of some inflation models of the 
universe \cite{kofman}. Parametrically excited systems are frequently nonlinear: even the parametrically driven pendulum is governed by the linear Mathieu equation only for small 
oscillations about its equilibrium positions. Nonetheless, valuable qualitative
insight into these systems may be gained by investigating their behavior in the vicinity of equilibrium points.

In the realm of ultracold quantum gases, 
studies of parametric resonance have featured Faraday patterns 
\cite{engels,nicolin3,longhi,nicolin1,capuzzi,nicolin2,balaz,balazneu}, Kelvin waves of quantized vortex lines \cite{simula}, 
self-trapped condensates \cite{ueda1,malomed1}, bright and vortex solitons \cite{ueda2,adhikari,mueller1}, and self-damping at zero temperature \cite{kagan}. Other investigations focus on the phenomenon of parametric resonance in 
the context of lower-dimensional Bose and Fermi gases \cite{graf,mueller2,mueller3}.
Parametric resonances have also been studied for optical lattices when the intensity of the lattice is periodically modulated in
time \cite{dalfovo} or when the lattice is shaken. In the latter case it was
even shown that a periodic driving can induce a quantum phase transition from a Mott insulator to a superfluid 
\cite{holthaus,arimondo}, paving the way for new techniques to engineer exotic phases \cite{sengstock1,sengstock2}.
Other novel experimental techniques \cite{malomed2,bagnato,ramos} to excite a Bose-Einstein condensate (BEC) of $^7$Li in the vicinity of a broad Feshbach resonance \cite{hulet}
by harmonic modulation of the $s$-wave scattering length, in contrast to the usual method of excitation 
by modulation of the trapping potential \cite{jin1,ketterle,shlyapnikov,jin2,torres,abdullaev}, have inspired investigations of parametric resonance and other phenomena in 
Refs.~\cite{adhikari,vidanovic,rapp,hamid}.
In the following we perform a systematic proof-of-concept study of the simplest case of parametric resonance in a  three-dimensional
BEC in a harmonic trap. To this end 
we demonstrate that within both a linear analytic and a nonlinear numeric analysis 
the stability characteristics of its equilibrium configurations can be changed by a periodic modulation of the attractive interaction.

\section{Variational Approach}

We start with modeling the dynamics of a BEC at zero temperature using the mean-field Gross-Pitaevskii (GP)  Lagrangian 
\begin{eqnarray}
L(t) = \int \bigg[ \frac{i \hbar}{2}\left(\psi \frac{\partial
\psi^*}{\partial t}- \psi^*\frac{\partial \psi}{\partial t} \right)
-\frac{\hbar^2}{2m}|\nabla \psi|^2  
- V({\bf r}) |\psi|^2  - \frac{2\pi\hbar^2 a(t)}{m} |\psi|^4 \bigg] \,d{\bf r}.
\label{lagrangian}
\end{eqnarray}
Extremizing the Lagrangian results in the well-known GP equation for the dynamics of the mean-field condensate wave function $\psi=\psi({\bf r},t)$.
In experiments generically a cylindrically-symmetric harmonic trapping potential $V({\bf r}) = m\omega_\rho^2(\rho^2+\lambda^2 z^2)/2$ is used, whose elongation is described by the 
trap anisotropy parameter $\lambda = \omega_z/\omega_\rho$. Furthermore, we assume that the $s$-wave scattering length is periodically modulated according to
\be
a(t) = a_0 + a_1 \sin{ \Omega t}.
\ee

In the following we will investigate how the stability of condensate equilibria depends on both the driving amplitude $a_1$ and the driving frequency $\Omega$, provided the 
time-averaged $s$-wave scattering length $a_0$ is slightly negative. In principle, this could be analyzed by solving the underlying GP equation for the condensate wave function 
$\psi=\psi({\bf r},t)$, which follows from extremizing the Lagrangian (\ref{lagrangian}). However, the thorough numerical analysis in Ref.~\cite{vidanovic} demonstrated convincingly 
that the dynamics of the GP equation can be well-approximated within a Gaussian variational ansatz for the GP condensate
wave function \cite{zoller1,zoller2}. Even for long evolution times and 
in the vicinity of resonances, where oscillations of the condensate are quite large, it was possible to reduce the GP partial differential equation to a set of ordinary differential 
equations for the variational parameters. 

Therefore, we follow the latter approach and employ the Gaussian ansatz
\begin{eqnarray}
\label{ansatz}
	\psi^{\rm G}(\rho,z,t) &=& {\mathcal N}(t)
	\exp\left[ -\frac{\rho^2}{2\tilde u_\rho^2} 
	+ i \rho^2 \phi_\rho \right] 
\exp\left[-\frac{z^2}{2\tilde u_z^2} 
	+ i z^2 \phi_z \right],
\end{eqnarray}
with time-dependent variational widths $\tilde u_{\rho}$, $\tilde u_{z}$, phases $\phi_{\rho}$, $\phi_{z}$, and normalization 
$
{\mathcal N}(t) = N^{1/2} \pi^{3/2} {{\tilde u}_\rho}^{-1} {{\tilde u}_z}^{-1/2}.
$
Inserting the Gaussian ansatz (\ref{ansatz}) into the GP Lagrangian (\ref{lagrangian}) and extremizing with respect to all variational parameters, we obtain at first explicit expressions 
for the phases $\phi_{\rho,z} = m \dot {\tilde u}_{\rho,z}/(2\hbar {\tilde u}_{\rho,z})$. We define the dimensionless time $\tau=\omega_\rho t$ and scale the variational widths by 
$u_{\rho,z} = {\tilde u}_{\rho,z}/a_{\rm ho}$, where $a_{\rm ho}=\sqrt{\hbar/(m\omega_\rho)}$ is the harmonic oscillator length.
Finally, we write the dimensionless driving function $p(\tau) = p_0 + p_1 \sin{\left( \Omega \tau/\omega_\rho \right)}$ according to the definitions 
$p_{0,1} = \sqrt{2/\pi} N a_{0,1}/a_{\rm ho}$.
The resulting dynamics for the widths $u_\rho$ and $u_z$ is then determined by a pair of coupled nonlinear ordinary differential equations:
\begin{eqnarray}
	\ddot{u}_{\rho} 
	+ u_{\rho}
	&=& \frac{1}{u_{\rho}^3}
	+ \frac{p(\tau)}{u_{\rho}^3u_z}, \nonumber \\
	\label{cyleqn1}
	\ddot{u}_z
	+ \lambda^2 u_z 
	&=& \frac{1}{u_z^3} 
	+ \frac{p(\tau)}{u_{\rho}^2u_z^2}.
\end{eqnarray}

For attractive two-body interactions, there is a critical value of the time-averaged interaction strength $p_0^{\rm crit}(\lambda)<0$ beyond which no equilibria exist
in the absence of parametric driving. This means physically that for $p_0^{\rm crit}(\lambda)<0$ the BEC always collapses.
The dependence of  $p_0^{\rm crit}(\lambda)<0$ 
on the trap anisotropy $\lambda$ must be evaluated numerically, as for example  in Ref.~\cite{hamid}. For $p_0^{\rm crit}(\lambda) <p_0<0$, there exists a pair of 
equilibrium points for Eqs.~(\ref{cyleqn1}), one stable and one unstable \cite{zoller1,zoller2}, which we denote with ${\bf u}_{0+}$ and ${\bf u}_{0-}$, 
respectively. We remark that the stability of these points in the absence of parametric driving is determined by evaluating the frequencies of collective modes for small oscillations 
about equilibrium \cite{hamid}. Whereas the equilibrium ${\bf u}_{0+}$ has real frequencies for all modes and is stable,
the equilibrium ${\bf u}_{0-}$ possesses an imaginary frequency for one mode, implying exponential behaviour and thus instability. 

\section{Linear Stability Analysis}
In view of a linear stability analysis we assume small oscillations about an equilibrium, write $u_\rho \approx u_{\rho0} + \delta u_{\rho}$ and $u_z \approx u_{z0} + \delta u_{\rho}$, 
and expand the nonlinear terms of Eqs.~(\ref{cyleqn1}) to first order in $\delta u_\rho$ and $\delta u_z$. We scale and translate time as $2t' + \pi/2= \Omega\tau/\omega_\rho$, define displacement and forcing vectors ${\bf x}(t')$ and ${\bf f}$,
\begin{eqnarray}
	{\bf x}(t') = \left( \begin {array}{c} \delta u_{\rho}(\tau) \\
	\delta u_{z}(\tau)\end {array} \right), \quad
	{\bf f} = 4\left(\frac{\omega_\rho}{\Omega}\right)^2
	\left( \begin {array}{c} \frac {p_1}{ u_{\rho0}^3 u_{z0}}\\
	{\frac {p_1}{ u_{\rho0}^{2} u_{z0}^2}}\end {array} \right),
\end{eqnarray}
and finally we introduce the matrices ${\bf A}$ and ${\bf Q}$ corresponding to coefficients of constant and periodic terms, respectively:
\begin{eqnarray}
	{\bf A} &=& 4\left(\frac{\omega_\rho}{\Omega}\right)^2
	\left( \begin {array}{cc} 4
	& \frac {p_0}{u_{\rho0}^3 u_{z0}^2}\\
	\frac {2p_0}{u_{\rho0}^3 u_{z0}^2}
	&  3\lambda^{2}+\frac{1}{u_{z0}^4}
	\end {array} \right), \nonumber \\
	{\bf Q} &=& -2\left(\frac{\omega_\rho}{\Omega}\right)^2 
	\left( \begin {array}{cc} 
	{\frac {3}{ u_{\rho0}^4{ u_{z0}}}}
	&  {\frac {1}{ u_{\rho0}^3 u_{z0}^2}} \\
	{\frac {2}{ u_{\rho0}^{3} u_{z0}^2}}
	& {\frac {2}{ u_{\rho0}^{2} u_{z0}^3}}
	\end {array}\right).
\end{eqnarray}
The result consists of two coupled inhomogeneous Mathieu equations
\be
	\ddot{\bf x}(t') + \left( {\bf A} - 2p_1{\bf Q}\cos{2t'} \right)  {\bf x}(t') 
	= {\bf f}\cos{2t}',
	\label{coupledmathieu}
\ee
whose solutions determine whether the underlying equilibrium is stable or unstable. 

The Mathieu equation, a special case of Hill's differential equation \cite{stegun}, has been studied extensively in the literature \cite{mclachlan}. Approaches to obtaining its
stability diagram include continued fractions \cite{stegun,risken,simmendinger}, perturbative methods \cite{nayfeh,younesian,mahmoud}, and infinite determinant methods 
\cite{lindh,pedersen,hansen,landa}. The problem has been treated in detail in Ref.~\cite{ozakin} for the study of the Paul trap, the stability of which is governed exactly by 
a set of coupled homogeneous Mathieu equations. Of importance to our particular problem are Refs.~\cite{kotowski,slane}, where it was shown that for both single and coupled Mathieu 
equations, a harmonic inhomogeneous term does not affect the location of stability borders to Eq.~(\ref{coupledmathieu}). 

In many approaches (see, for example, Ref.~\cite{landa}), Eq.~(\ref{coupledmathieu}) is reformulated as a first-order non-autonomous Floquet system 
\be
\label{1storder}
	\dot \phi = {\bf G}(t') \phi, 
\ee
where $\phi = \left[{\bf x} \;\; {\bf \dot x} \right]^T$, and
\be
	{\bf G}(t') = \left( \begin{array}{cc}
	{\bf 0} & {\bf 1}_{2} \\
	-\left( {\bf A} - 2p_1{\bf Q}\cos{2t'} \right) & {\bf 0}
	\end{array} \right)
\ee
is $\pi$-periodic.
Linearly independent solutions to the Floquet problem may be written as a fundamental matrix solution $\Phi(t')$ with 
the initial condition $\Phi(0) = {\bf 1}_4$:
\be
\label{fundmat}
	\dot \Phi(t') = {\bf G}(t')\,\Phi(t').
\ee
It can be shown that solutions to Eq.~(\ref{fundmat}) are stable if the eigenvalues $\lambda_n$ of $\Phi(t'=\pi)$ satisfy $|\lambda_n| \leq 1$. Further, we define characteristic exponents $\beta_n$ such that $\lambda_n = e^{\beta_n \pi}$, with the result that stable solutions ${\bf x}_{1,2}(t')$ to Eq.~(\ref{coupledmathieu}) may be written in the form \cite{yakubovich}
\be
{\bf x}_{1,2}(t') = e^{\pm\beta t'} \sum_{n=-\infty}^{\infty} {\bf b}_{2n} e^{2in t'}.
\label{floquet}
\ee
On the stability borders, Eq.~(\ref{floquet}) provides one linearly independent solution that is periodic; a second is non-periodic and grows linearly with time. In constructing the linear stability diagram, we use two complementary approaches: a matrix continued fraction approach based on Refs.~\cite{risken,simmendinger} determines the stability borders analytically, while a numerical integration of Eq.~(\ref{fundmat}) to obtain the characteristic multipliers $\lambda_n$ determines the stable or unstable character of the respective diagram regions.

\begin{figure}[t]
\begin{center}
	\includegraphics{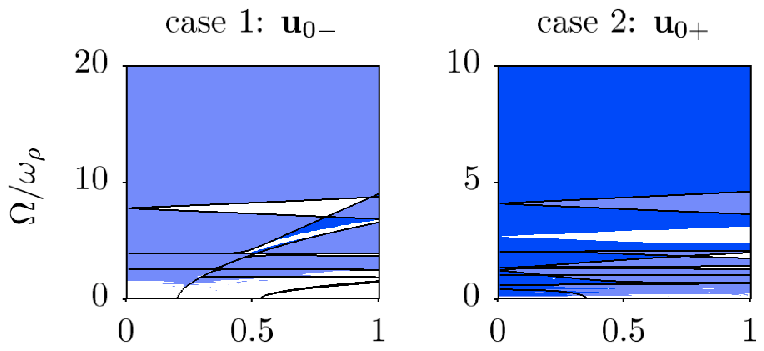}\\
	\vspace{-13.7mm}\hspace{-78mm}{\scriptsize(a)}\vspace{7mm}\\
	\includegraphics{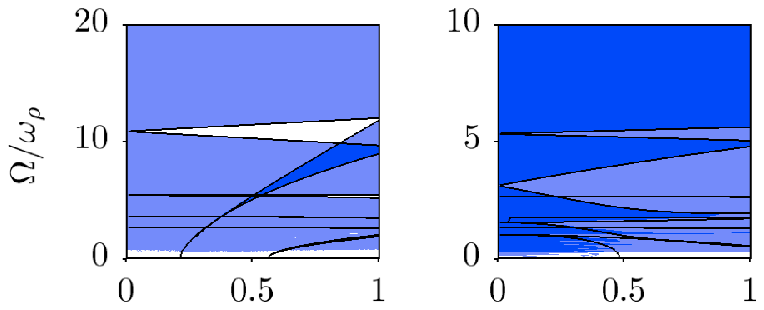}\\
	\vspace{-13.7mm}\hspace{-78mm}{\scriptsize(b)}\vspace{7mm}\\
	\includegraphics{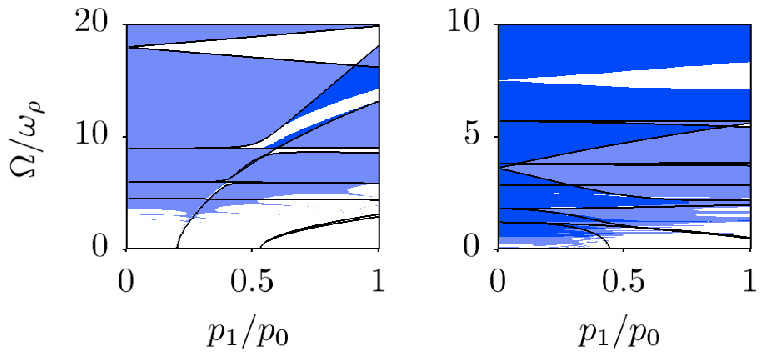}\\
	\vspace{-14.85mm}\hspace{-78mm}{\scriptsize(c)}\vspace{8mm}\\
\caption{Linear stability diagrams from solving two coupled Mathieu equations
for the unstable (case 1) and stable (case 2) equilibrium positions of a cylindrically-symmetric BEC for three 
values of the trap anisotropy $\lambda$: (a) $\lambda=0.2$ (a cigar-shaped BEC), (b) $\lambda=1$ (spherical BEC), and (c) $\lambda=2.6$ (pancake-shaped BEC). White regions correspond to unstable 
solutions, darkest shaded regions to stable solutions, and lightly shaded regions correspond to marginally stable solutions -- regions where only one of two available collective oscillation 
modes is stable.}
\label{comicmat}
\end{center}
\end{figure}

By substitution of the Floquet ansatz (\ref{floquet}) into Eq.~(\ref{coupledmathieu}), we obtain a third-order recurrence relation for the Fourier coefficients ${\bf b}_{2n}$:
\be
	\label{recursion2}
	\left[ {\bf A} + \left( \beta + 2in \right)^2  {\bf I} \right] {\bf b}_{2n} 
	- p_1{\bf Q} \left( {\bf b}_{2n+2} + {\bf b}_{2n-2} \right) = \bf 0.
\ee
We define the ladder operators ${\bf S}_{2n}^{\pm} {\bf b}_{2n} = {\bf b}_{2n\pm2}$ as
\be
\hspace*{-0.1cm}	{\bf S}_{2n}^{\pm} = \left\{ {\bf A} + \left[ \beta + 2i\left(n+1\right) 
	\right]^2{\bf I} - p_1{\bf Q}{\bf S}_{2n\pm2}^{\pm} \right\}^{-1} p_1  {\bf Q},
\ee
and by repeated re-substitution of these ladder operators into the recursion relation (\ref{recursion2}) for $n$ increasing and decreasing from zero, we obtain a tri-diagonal 
matrix-valued continued fraction relating the parameters ${\bf A}$, ${\bf Q}$, and $\beta$:
\begin{eqnarray}
{\bf\bigg(} {\bf A}+\beta^2{\bf I} - p_1^2{\bf Q} 
	\bigg\{ \left[{\bf A}	+ \left(\beta + 2i \right)^2{\bf I} - \dots \right]^{-1}
+ \left[ {\bf A} + \left( \beta - 2i \right)^2{\bf I} - \dots \right]^{-1} \bigg\}
	{\bf Q} {\bf\bigg)} 	{\bf b}_{0} 	= \bf 0.
	\label{matrixinversion}
\end{eqnarray}
In order to obtain a non-trivial solution for ${\bf b}_0$, the determinant of the matrix-valued continued fraction (\ref{matrixinversion}) must vanish. Since on the stability borders one linearly independent solution is periodic, it suffices to set $\beta = \{ 0,\pm i\}$ in Eq.~(\ref{matrixinversion}) and truncate the continued matrix inversion, determining the stability borders in terms of the dimensionless driving amplitude $p_1$ and the driving frequency $\Omega$. Empirically, a truncation at ${\bf S}_4^{\pm}$ proves to be sufficient for accurate results.
The resulting stability 
for particular values of $\Omega$ and $p_1$ is then shown in the linear stability diagram of Fig.~\ref{comicmat} for three 
characteristic values of the trap anisotropy $\lambda$. Our results indicate that for the unstable (stable) equilibrium position, the largest region of stability (instability) occurs 
for a pancake-shaped BEC, i.e., for $\lambda>1$.

\begin{figure}[t]
\begin{center}
	\includegraphics{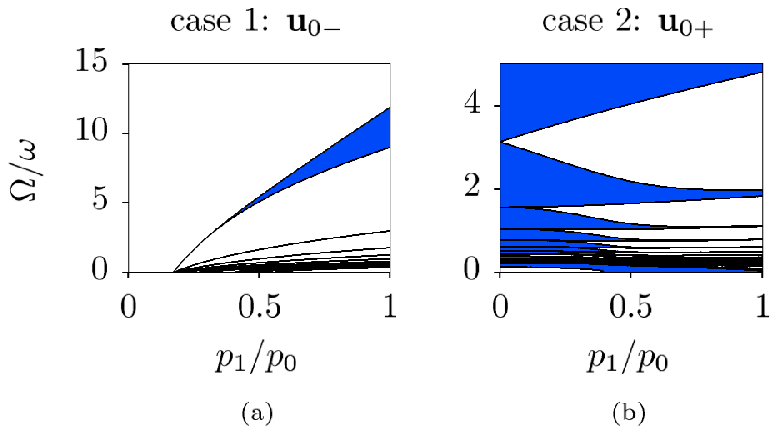}
\caption{Linear stability diagrams from solving a single Mathieu equation
for (a) unstable and (b) stable equilibria of a 3D spherically symmetric BEC. Shaded and white regions indicate stable and unstable solutions, respectively. The presence 
of a stable region in (a) indicates that an originally unstable equilibrium might be stabilized by parametric excitation.}
\label{iso}
\end{center}
\end{figure}

As a special case the stability borders for the isotropic condensate follow from the 3D spherically symmetric version of Eq.~(\ref{ansatz}) and
are given by a separate
continued fraction, obtained by an analogous process for a single inhomogeneous Mathieu equation. The 
resulting stability diagrams are shown in Fig.~\ref{iso}. The isotropic case allows
a direct analogy between the condensate and the parametrically driven pendulum: the pendulum too is described by a single Mathieu equation, however the inhomogeneous 
term in the equation of motion for the BEC corresponds to a direct periodic driving in phase with the parametric driving. It was shown in Ref.~\cite{kotowski} that a periodic inhomogeneity 
has no effect on the stability borders for a single Mathieu equation, so Fig.~\ref{iso} is simply a transformation of the standard Ince-Strutt stability diagram for the parametrically driven 
pendulum \cite{stegun}, for the relevant experimental parameters of the BEC.

While the results of linear stability analysis for the coupled Mathieu system are qualitatively similar to those for the single equation, there are a number of notable changes between 
Figs.~\ref{comicmat} and \ref{iso}. First, for the coupled Mathieu equations, there exist a set of stability regions not attainable by the analytic method used here. These are displayed 
without black borders in Fig.~\ref{comicmat}, and correspond to the so-called ``combined resonances" of the system \cite{hansen,ozakin}. These regions are attainable by numerical stability 
analysis of the Mathieu equations \cite{yakubovich,ozakin,syranian}, which was used to generate the colored background regions of Fig.~\ref{comicmat}. It is notable in Fig.~\ref{comicmat} that 
these anomalous regions are not present for $\lambda=1$.

A second and important difference from the single to the coupled Mathieu equations is the appearance of a new region, shaded white and issuing from $\Omega/\omega_\rho\approx10$ in 
Fig.~\ref{comicmat} for $\lambda=1$, case 1. This region is identified with the instability of the quadrupole collective mode, which does not appear in a one-dimensional analysis. This result 
implies that a three-dimensional analysis might result in further changes to the linear stability diagram of Fig.~\ref{comicmat} for $\lambda=1$, case 1.

\section{Nonlinear Numerics}
\begin{figure}[t]
\begin{center}
	\includegraphics{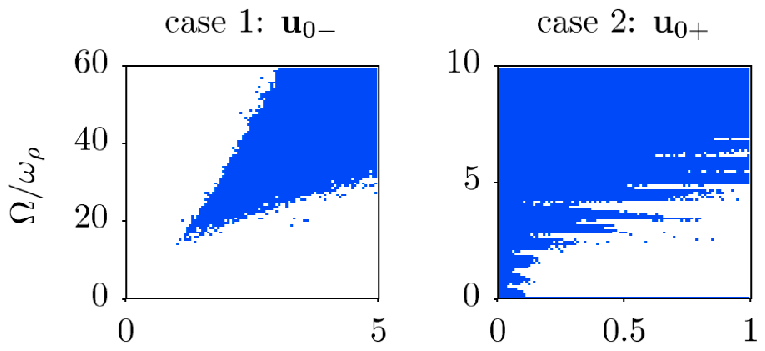}\\
	\vspace{-13.7mm}\hspace{-78mm}{\scriptsize(a)}\vspace{7mm}\\
	\includegraphics{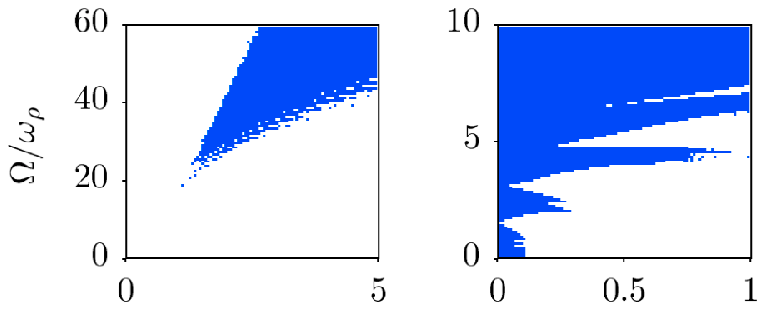}\\
	\vspace{-13.7mm}\hspace{-78mm}{\scriptsize(b)}\vspace{7mm}\\
	\includegraphics{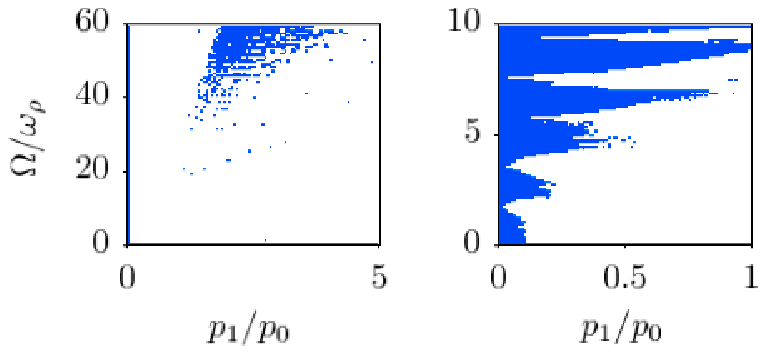}\\
	\vspace{-15.2mm}\hspace{-78mm}{\scriptsize(c)}\vspace{7mm}\\
\caption{Nonlinear stability diagrams for the unstable (left) and stable (right) equilibria of a cylindrically-symmetric BEC, for three values of the trap anisotropy $\lambda$: 
(a) $\lambda=0.2$ (a cigar-shaped BEC), (b) $\lambda=1$ (spherical BEC), and (c) $\lambda=2.6$ (pancake-shaped BEC).  
Shaded and white regions indicate stable and unstable solutions, respectively.}
\label{comicnonlin}
\end{center}
\end{figure}

As the underlying equations of motion (\ref{cyleqn1}) are inherently nonlinear, a linear analytic stability analysis alone is not sufficient to fully
investigate the phenomenon of parametric
resonance. Therefore, we have also performed a detailed numerical stability analysis by integrating the equations of 
motion (\ref{cyleqn1})
over a long time-period using a Runge-Kutta-Verner 8(9) order algorithm, incrementing through pairs $(p_1,\Omega)$ and recording divergent solutions to obtain the corresponding 
stability diagram. The corresponding results are shown in Fig.~\ref{comicnonlin} for the same three values of the trap anisotropy $\lambda$ as in Fig.~\ref{comicmat}.

The results for the originally stable equilibrium ${\bf u}_{0+}$ show both qualitative and even quantitative similarity to the linear stability analysis of Fig.~\ref{comicmat}, i.e., a similar 
tonguelike structure of unstable regions issuing from certain points on the vertical axis. The originally unstable equilibrium ${\bf u}_{0-}$ also shows qualitative similarity to the 
linear analysis of Fig.~\ref{comicmat}, however the region of stability begins only for much larger modulation frequency $\Omega$ and dimensionless driving amplitude $p_1$. These results 
are both reasonable, as the linear stability analysis is only valid for small oscillations -- corresponding to large $(p_1,\Omega)$ for equilibrium ${\bf u}_{0+}$ and small $(p_1,\Omega)$ 
for equilibrium ${\bf u}_{0-}$. Furthermore, in contrast to Fig.~\ref{comicmat}, we find in the nonlinear stability diagram of Fig.~\ref{comicnonlin} that stability is more easily achieved 
for a cigar-shaped BEC, i.e., for $\lambda<1$.

Further comparison of the linear and nonlinear stability diagrams of Figs.~\ref{comicmat} and \ref{comicnonlin} shows the possibility of both simultaneous stability of the equilibrium 
positions, and even the possibility of a complete reversal of the stability characteristics. In the latter case, the smaller equilibrium position would become the only stable width of 
the condensate, which should be experimentally observable. A final observation, applicable to the originally unstable equilibrium in both linear and nonlinear cases, is the existence of 
a minimum driving amplitude $p_1^{\rm min}$ necessary to stabilize the condensate. The value $p_1^{\rm min}=u_0(5u_0^4-1)\approx0.17~p_0$ 
is exactly attainable for the linear analysis of the isotropic 
condensate, and in Fig.~\ref{comicmat} is approximately independent of $\lambda$ in the considered
range [0.2,2.6]. For the nonlinear analysis, $p_1^{\rm min} \approx1.2~p_0$ is also approximately 
independent of $\lambda$. This feature will have implications for an experiment, as in conjunction with the width of the Feshbach resonance, it dictates the minimum modulation of the applied 
magnetic field necessary to stabilize the condensate.

We note that any stability diagram depends on the particular choice for the time-averaged dimensionless interaction strength $p_0$. The concrete results in 
Figs.~\ref{comicmat}--\ref{comicnonlin} were obtained for the particular value $p_0=0.9~p_0^{\rm crit}(\lambda)$. As $p_0$ approaches $p_0^{\rm crit}(\lambda)$, we generically observe 
a growth of the stable (unstable) regions in case 1 (2).

\section{Conclusions}

Finally, we conclude that
our proof-of-concept investigation has unambiguously shown that the phenomenon of parametric resonance should also 
be experimentally observable in terms of changed stabilities for the equilibrium configurations of a three-dimensional
BEC in a harmonic trap with a periodic modulation of the attractive interaction.
However, due to the intrinsic nonlinear nature of the underlying GP mean-field theory, a linear analysis, like in the Paul trap, is not sufficient to quantitatively study the stability 
diagram. Thus, in order to achieve a destabilization (stabilization) of a stable (unstable) BEC equilibrium in an experiment, a corresponding numerical nonlinear analysis is indispensable. 
Regardless, a linear stability analysis provides an intuitive and qualitative understanding of the physics of parametric resonance in BECs.

In the present letter we have focused our attention upon a periodic modulation of the $s$-wave scattering length around a slightly negative value, which restricts the number of particles 
in a BEC to the order of a few thousand \cite{bradley,bradleyerrata}. However, the phenomenon of parametric resonance might be more important for dipolar BECs, where in addition to a 
repulsive short-range and isotropic interaction, also a long-range and anisotropic dipolar interaction between atomic magnetic or molecular dipoles is present. Provided that the dipolar 
interaction is smaller than the contact interaction, a stable dipolar BEC does exist. But a larger dipolar strength leads to mutual existence of both a stable and an unstable dipolar 
BEC \cite{eberlein} whose stability might be changed via a periodic modulation of the harmonic trap frequencies or the $s$-wave scattering length. In that context it might also be of 
interest to estimate how quantum fluctuations, which are non-negligible for a larger dipolar interaction strength \cite{lima1,lima2}, change the stability diagram. The case for dipolar Fermi 
gases is probably even more interesting from the point of view of parametric resonance, as for any dipolar strength a stable equilibrium coexists with an unstable one \cite{lima3,lima4}. Thus, 
parametric resonance may offer a simple efficient approach for realizing equilibria of dipolar quantum gases whose properties have so far not yet been explored.

\section*{Acknowledgement}
We are grateful to Jochen Br\"uggemann for discussions and 
Antun Bala\v{z} and Haggai Landa for their valuable input. This project was supported by the DAAD (German Academic Exchange Service) within the program RISE (Research Internships in Science and Engineering).\\
	
\section*{References}

\end{document}